\documentclass[12pt]{article}
\usepackage{epsf,citesort,subfigure}
\newcommand{\cd}{\makebox[0.08cm]{$\cdot$}}

\title{Deuteron electromagnetic form factors in the Light-Front Dynamics}
\author{J. Carbonell\thanks{e-mail: carbonel@isn.in2p3.fr}
\\ 
{\small \em  Institut des  Sciences Nucl\'{e}aires, 
53 avenue des Martyrs, 38026 Grenoble Cedex, France}
\and
V.A. Karmanov\thanks{e-mail: karmanov@sci.lebedev.ru}
\\ 
{\small \em Lebedev Physical Institute, Leninsky Prospekt 53, 117924 
Moscow, Russia} }

\begin{document}

\maketitle
\bibliographystyle{unsrt}

\begin{abstract}

{The deuteron form factors are calculated in the framework of the
relativistic nucleon-meson dynamics, by means of the explicitly covariant light-front approach. 
The inflluence of the nucleon 
electromagnetic form factors is discussed. At $Q^2\leq 3$ (GeV/c)$^2$ the prediction for the structure 
function $A(Q^2)$ and for the tensor polarization observable $t_{20}$ are in agreement with the recent data of CEBAF/TJNAF.} 
\end{abstract}

\section{Introduction}\label{intro}

Recently the first experimental data obtained at Thomas Jefferson  National
Accelerator Facility (former CEBAF) on the deuteron structure function $A$ and
tensor polarization $t_{20}$, measured  in the elastic $ed$ scattering, have
been reported \cite{Kox,Gomez,Furget_98,BB_98}. They are rather precise and
correspond to a maximum momentum transfer of 6 (GeV/c)$^2$.  Since in the
$ed$ collision the deuteron gets from the electron a momentum comparable  with the
deuteron mass, these data are probing the trully relativistic dynamics inside the deuteron.

Our preliminary calculations of the form factors and of $t_{20}$ have been 
presented in~\cite{cdkm}. We present in this paper the details of our approach,
discuss the influence of the different choices of nucleon electromagnetic form factors
parametrization and compare our results with the reported experimental data.

In our calculations we assume the deuteron structure to be determined by the
relativistic  nucleon-meson dynamics. Namely, we suppose that the nucleons in
the deuteron interact by exchanging relativistic mesons. We take the
same set of mesons and parameter values used in the construction of the Bonn
potential~\cite{bonn}. However, we do not make any nonrelativistic potential
approximation  and calculate  the one-boson-exchange kernel in relativistic
form, as it appears from field theory.

In the momentum tranfer region scanned in TJNAF, the relativistic effects related to the
nucleons motion, to the spin
rotations and to the retardation of the exchanged mesons, should manifest themselves
in full measure and be of crucial importance in describing the data.
Therefore we believe that their main part can be properly
taken into account even in an approximate relativistic calculation. On the other hand, it is
important to work in an approach which provides a clear physical
interpretation of the incorporated effects.

The deuteron form factors are calculated  in the framework of the  explicitly
covariant version of the Light-Front Dynamics (LFD) recently reviewed in
\cite{cdkm}.  In this approach, the state vector  is defined 
on the light-front plane of general  position $\omega\cd x =0$, where
$\omega$ is a four-vector such as  $\omega^2=0$.  This restores the relativistic
covariance lost in the standard  light-front approach which is obtained as a
particular case for $\omega=(1,0,0,-1)$. The relativistic wave functions -- the
Fock components of the state vector --  are the closest couterparts of the
nonrelativistic ones. This allows one to benefit from the knowledge of
the nonrelativistic wave functions and to incorporate selfconsistently the
relativistic effects.  Using the light-front plane 
leads to  significant simplifications due to the absence of 
vacuum  fluctuations.  The diagram technique is a three-dimensional one and 
can be interpreted in terms of time-ordered physical processes.  Therefore the selection of the
diagrams contributing to the kernel or to the form factors, whose  numerical
estimation is rather difficult in any approach, can be supported by the intuitive
understanding of the corresponding physical  process.  All that allows, in a
given relativistic dynamics, to carry out the calculations on a satisfactory
level of confidence. 

We mention also few other approaches to calculate the relativistic wave 
function and the deuteron electromagnetic  form factors. 
In the standard version of LFD, defined on the plane $t+z=0$, these form 
factors  were calculated in \cite{GK_84,CCKP_88,FGKS_89,FFS_93}. 
The results obtained in \cite{ztjonall,HT_89} are based on the Bethe-Salpeter approach and the works
\cite{ACG_80,VDG_95} use a three-dimensional reduction of the Bethe-Salpeter
function -- the Gross wave function \cite{gross2}. 

In sect. \ref{wf} we present the relativistic deuteron wave function on the
light front. In sect. \ref{emv} the electromagnetic vertex of the deuteron in
the impulse approximation, based on the one-body electromagnetic nucleon current, is
discussed. In sect. \ref{contact} we discuss the contact (instantaneous) 
interaction, corresponding to the $NNB\gamma$ vertex  ($B$ is an exchanged
meson) which is a correction to the one-body current. In sect. \ref{ffs} we
explain how to extract the form  factors, separating them from the nonphysical
contributions. Section \ref{sec_nemff} is devoted to the evaluation of 
the nucleon electromagnetic form factors influence in the deuteron observables.
Section \ref{numer} contains a comparison of our results with the experimental data
and concluding remarks.

\section{Wave function}\label{wf}

The wave functions are the Fock components of the state vector defined  on the
light-front plane $\omega\cd x =0$.  The explicit covariance  allows one to
construct the general form of the light-front wave  function for a system with a
given spin.  The relativistic deuteron wave  function on the light front
contains six spin components, in contrast  to two components -- S and D-waves
-- in the nonrelativistic case. Its  general form is given in
\cite{karm81,ck-deut}. Below we will keep only  three dominating components:  
\begin{eqnarray}\label{eq1} 
{\mit\Psi}_{\lambda}^{\sigma_2\sigma_1}&=&
\sqrt{m}e_{\lambda}^{\nu}(p) 
\bar{u}^{\sigma_2}(k_2)\phi_{\nu}U_c\bar{u}^{\sigma_1}(k_1), 
\nonumber\\
\phi_{\nu}&=& 
\varphi_1\frac{(k_1- k_2)_{\nu}}{2m^2}
+\varphi_2\frac{1}{m}\gamma_{\nu}                           
-\varphi_5\frac{i}{m^2\omega\cd p}\gamma_5
\epsilon_{\nu\alpha\beta\gamma}          
k_1^{\alpha}k_2^{\beta}\omega^{\gamma},
\end{eqnarray} 
where $p$ and $k_{1,2}$ are the on mass shell deuteron and the nucleon momenta, 
$e^{\lambda}_{\nu}(p)$ is the deuteron polarization vector, 
$\bar{u}^{\sigma}(k)$ is the nucleon spinor, $U_c$ is the charge  conjugation
matrix. We notice that the wave function defined on the light-front plane
depends on the orientation of this plane through the argument $\omega$. 

In the system of reference where $\vec{k}_1+\vec{k}_2=0$ the function
(\ref{eq1}) obtains the more transparent form:
\begin{eqnarray}\label{eq2} 
{\mit\Psi}^{\lambda}_{\sigma_2\sigma_1}(\vec{k},\vec{n})& =&
\sqrt{m}w^\dagger_{\sigma_2} \psi^{\lambda}(\vec{k},\vec{n})\sigma_y 
w^\dagger_{\sigma_1}\ , 
\nonumber\\
\vec{\psi}(\vec{k},\vec{n}) & = & f_1\frac{1}{\sqrt{2}}\vec{\sigma} + 
f_2\frac{1}{2}(\frac{3\vec{k}(\vec{k}\cd\vec{\sigma})}{\vec{k}^2} 
-\vec{\sigma})
+f_5\sqrt{\frac{3}{2}}\frac{i}{k}[\vec{k}\times \vec{n}],
\end{eqnarray} 
where $\vec{k}$ is the value of $\vec{k}_1$ in this system of  reference,
$\vec{n}$ is the direction of $\vec{\omega}$ in this system,  $w_{\sigma}$ is
the two-component nucleon spinor. The scalar functions  $f_{i}$ depend on the
scalars $k \equiv \vert \vec{k} \vert $ and  $z=\vec{n}\cd\vec{k}/k$.  One can
find a special  representation \cite{ck-deut} in which the wave function
(\ref{eq1})  obtains the form (\ref{eq2}) in arbitrary system of reference. 
The  components $\varphi_i$ in (\ref{eq1}) and $f_i$ in (\ref{eq2}) are 
linearly related with each other \cite{cdkm,ck-deut}.

The wave function (\ref{eq1}) was calculated in \cite{ck-deut} in a 
perturbative way incorporating the full relativistic one-boson exchange (OBE) kernel found in  LFD,
with the corresponding nonrelativistic wave function as zero order
approximation.  This calculation can be considered as the perturbation theory
developed in terms of the difference between the relativistic kernel and the
nonrelativistic potential. For the OBE kernel the set of mesons, coupling
constants and form factors corresponding to the Bonn model \cite{bonn} were
used. The solution thus obtained is approximate in two ways: first for its
perturbative character does not provide an exact solution of the LFD equation
and second because the parameters of the OBE kernel were kept to their original
values i.e. were not fitted in the LFD relativistic framework. The accuracy of our solution in reproducing
the low energy deuteron observables was estimated at the level of 20\% in norm.

In nonrelativistic region of $k$, components $f_1$ and $f_2$ found this
way turn into the usual S- and D-waves, whereas other components become
negligible. However, starting from $k\approx 0.5$  GeV/c, component $f_5$
dominates over all other components, including $f_1$ and $f_2$. 

The physical meaning of this dominating extra component has been  clarified
\cite{dkm95} by comparing, in $1/m$ approximation, the  analytical expression
for the amplitude of the deuteron  electrodisintegration near threshold with
the nonrelativistic one,  including meson exchange currents. For the isovector
transition, in the  region where the so called pair term with the pion exchange
dominates,  this component (together with a similar component in the scattering 
state) automatically incorporates 50\% of the pair  term contribution 
and therefore dominates too. Another 50\% is given by the contact interaction (see below).

\section{Electromagnetic vertex}\label{emv}

\begin{figure}[htbp]
\begin{center}
\epsfxsize=6cm
\centerline{\epsfbox{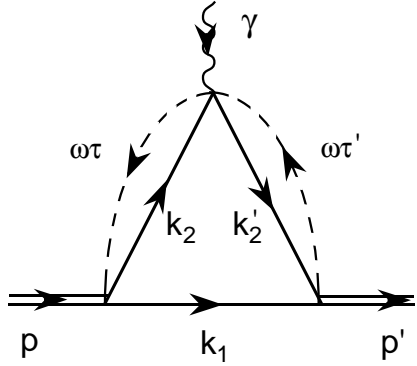}}
\caption{Electromagnetic vertex of the deuteron.}\label{f1_ks92_b}\end{center}
\end{figure}

The electromagnetic vertex $J_{\mu\nu}^{\rho}$ we use to calculate  the
deuteron form factors corresponds to the sum of the impulse  approximation (IA)
(figure \ref{f1_ks92_b}) and of the contact interaction (C) (figure \ref{ct4})
which takes partially into account the two-body current: 
\begin{equation}\label{eq4a}
J_{\mu\nu}^{\rho}=J_{\mu\nu}^{\rho}({\rm IA})+J_{\mu\nu}^{\rho}({\rm C}).
\end{equation}
These contributions are explained in what follows.

The deuteron IA electromagnetic vertex is shown in figure \ref{f1_ks92_b}. The
dashed line is associated with a  fictitious particle -- the so called spurion.
It reflects the fact that, although all  the momenta are on the corresponding
mass shells, the wave function is  off energy shell, and therefore there is no
any conservation law  between all the components of the nucleon and the
deuteron  four-momenta. The spurion momentum just absorbs their nonzero 
difference and one has $k_1+k_2-p=\omega\tau$. To avoid misunderstanding, we
emphasize that the spurion line does not imply the presence of an extra
particle in the intermediate state in figure \ref{f1_ks92_b} which contains two nucleons  only. 

The electromagnetic amplitude in the impulse approximation is
derived by applying the rules of the graph techniques \cite{cdkm} to  figure
\ref{f1_ks92_b}. It has the form:
\begin{eqnarray}\label{eq4}                                                   
&&\langle \lambda ' \vert J^\rho({\rm IA}) \vert \lambda \rangle
   =e^{*\mu}_{\lambda'}(p') J_{\mu\nu}^{\rho}({\rm IA})e^{\nu}_{\lambda}(p)\ ,\\              
&&J_{\mu\nu}^{\rho}({\rm IA})=\frac{m}{(2\pi)^3}\int  
Tr[ \phi'_{\mu} \; (\hat{k}'_2+m) \; \Gamma^{\rho} \; (\hat{k}_2+m)\;
\phi_{\nu}(\hat{k}_1-m) ]                              
\; \frac{d^3k_1}{(1-x)^2 2\varepsilon_{k_1}},\nonumber
\end{eqnarray}
where $x=\omega\cd k_1/\omega\cd p$. The wave function 
$\phi_{\nu}=\phi_{\nu}(k_1,k_2,p,\omega\tau)$ given by eq.(\ref{eq1})
corresponds to the deuteron initial state, while
$\phi'_{\mu}=\phi_{\mu}(k_1,k'_2,p',\omega\tau')$ corresponds to its final
state.   $\Gamma_{\rho}$ is the electromagnetic vertex of the nucleon:  

\begin{equation} \label{eq5} 
\Gamma^{\rho}=F_1\gamma^{\rho}  
+\frac{iF_2}{2m}\sigma^{\rho\alpha}q_{\alpha},
\end{equation}
$F_1$ and $F_2$ being the nucleon elecromagnetic form factors.

The expression for the trace in (\ref{eq4}) is obtained as follows. To each
nucleon line in figure~\ref{f1_ks92_b} we associate the LFD nucleon
propagator,  with the $NN\gamma$ vertex  we associate expression (\ref{eq5}), and
with $d-NN$ vertices the deuteron wave functions.
We get in this way the following product of spin  matrices:

\begin{equation}\label{ffe1}
e^{*\mu}_{\lambda'}(p')\left\{[\gamma_0\phi'_{\mu} U_c\gamma_0]^{\dagger}
(\hat{k}'_2+m)
\Gamma^{\rho}(\hat{k}_2+m)\phi_{\nu}U_c\right\}_{\beta\alpha}
 e^{\nu}_{\lambda}(p) (\hat{k}_1+m)_{\beta\alpha}.
\end{equation}

Factor $\{\ldots\}_{\beta\alpha}$ in (\ref{ffe1}) corresponds to the upper
line of the diagram and the factor $(\hat{k}_1+m)_{\beta\alpha}$
corresponds to the lower line. We attach the deuteron wave function to the
upper one. We keep in (\ref{ffe1}) the matrix indices $\alpha$ and
$\beta$ explicitly.  Since both nucleon lines are passed in the same direction, the order of
indices $\beta,\alpha$ is the same. This means that one of the factors in
(\ref{ffe1}) (we take the second one) is a transposed matrix. The factor
$[\gamma_0\phi'_{\mu}U_c\gamma_0]^{\dagger}$ originates from the conjugated
final deuteron wave function:  
\begin{equation}\label{wfc} \left(
\bar{u}(k'_2)\phi'_{\mu}U_c\bar{u}(k_1)\right)^{\dagger}=
u(k_1)[\gamma_0\phi'_{\mu}U_c\gamma_0]^{\dagger}u(k'_2) =  - u(k_1)
\bar{\phi}'_{\mu}  u(k'_2),                                         
\end{equation}                                                                            
where $\bar{\phi}=\gamma_0 \phi^{\dagger} \gamma_0$.  
With the wave function (\ref{eq1}) we get:
$$[\gamma_0\phi'_{\mu}U_c\gamma_0]^{\dagger}=-U_c\phi'_{\mu}$$ 
and, hence, obtain the factor: 
$$-U_c(\hat{k}_1+m)^tU_c=(\hat{k}_1-m),$$ 
that gives the trace in eq.(\ref{eq4}). A more detailed derivation of (\ref{eq4})
by means of the LFD graph technique can be found in \cite{cdkm}. Next section
is devoted to the derivation of the contact contributions.

\section{The contact interaction}\label{contact}

\begin{figure}[hbtp] 
\begin{center}
\epsfxsize=7cm\epsfysize=7cm\centerline{\epsfbox{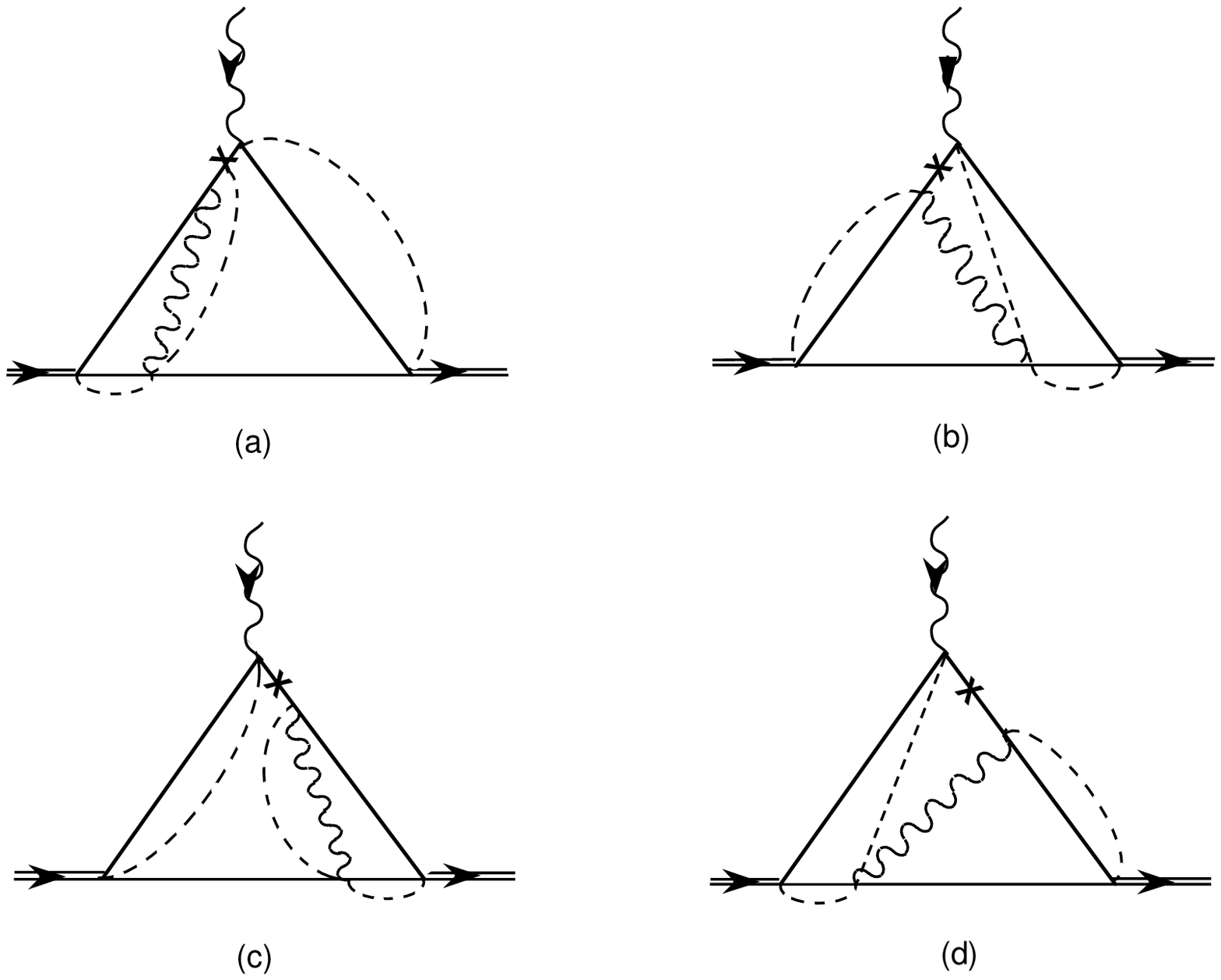}} \caption{Contact
term contributions to the electromagnetic interactions with deuteron.}\label{ct4}  
\end{center}
\end{figure}

A peculiarity of the LFD is the existence, in addition to the impulse 
approximation, of the so called instantaneous (or contact) interaction in the
$NNB\gamma$ vertex, where $B$ is any of the mesons $\pi,\rho,\ldots$, building
the NN potential.  Its contribution to the  electromagnetic vertex in $g^2$
order is shown in figure \ref{ct4}.  The cross on a fermion line means that
this line is not to be associated with a propagator, but with a factor
proportional to $\hat{\omega}=\omega_{\mu}\gamma^{\mu}$. In  the standard
approach this is the well known  instantaneous interaction vertex  $\gamma^{\dagger}$. The
nucleon electromagnetic vertex $\Gamma_{\rho}$  is determined, as usual, by the two
nucleon electromagnetic form factors given in eq.(\ref{eq5}).

In evaluating the contact amplitude, we have taken into account the sum over
all six  mesons contributing to the Bonn potential, with the parameters used in the
Bonn model \cite{bonn}. This amplitude incorporates partially the two-body currents.
In principle more complicated diagrams, of order higher than $g^2$, may also contribute.
Since our wave function takes into account corrections associated with the first degree of
$g^2$, we keep the same order for the electromagnetic vertex.  This means that
we consider only the diagrams of figure \ref{ct4}  and in calculating these
diagrams we omit component $f_5$ in the wave function, since its
contribution multiplied by the contact interaction is of higher order than $g^2$. 
To calculate the effects beyond the $g^2$ order, one has also to take into
account the higher order irreducible contributions to the OBE kernel.  
One can expect that these contributions are incorporated in this  kernel
on a phenomenological level by an appropriate fit of its parameters (coupling
constants, meson masses and cutoff parameters). However, this is not the
case for the electromagnetic vertex. 

The contact amplitude is given by the sum of two terms which differ
by the relative time order of the contact and the electromagnetic vertices:  
\begin{equation}\label{c0}                                                      
J^{\rho}_{\mu\nu}({\rm C})=J^{\rho}_{\mu\nu}({\rm C})\vert_{left} 
+ J^{\rho}_{\mu\nu}({\rm C})\vert_{right}.                                                 
\end{equation}                                                                  

The $J^{\rho}_{\mu\nu}({\rm C})\vert_{left}$ term corresponds to the sum of diagrams
(a) and (b) in figure \ref{ct4}, which differ from each other by the
relative time order of the meson vertices.
Both of them have the contact vertex on the left hand side of the electromagnetic one. 
The opposite order is assumed in $J^{\rho}_{\mu\nu}({\rm C})\vert_{right}$  
which corresponds to  diagrams (c) and (d) in figure \ref{ct4}. 
In addition one has to take the sum over all mesons contributing to the interaction.

For a given meson, the amplitude corresponding to the sum of diagrams (a) and (b) in
figure~\ref{ct4} -- left contact term -- has the form:                                                      

\begin{eqnarray}\label{c1}                                                      
J^{\rho}_{\mu\nu}\vert_{left}=m\int Tr\{ (-U_c\bar{\phi}'_{\mu})                                                         
(\hat{k}'_2+m)\Gamma^{\rho} \left(-\frac{\hat{\omega}} {2\omega \cd            
k'_2}\right)V (\hat{k}_2+m)\phi_{\nu}U_c \nonumber \\                                                                    
\times [(\hat{k}'_1+m) V(\hat{k}_1+m)]^{t} \} \frac{1}{(\mu^2+\vec{K}\,^2)} \frac{d^3k_1}                                 
{(2\pi)^3(1-x)2\varepsilon_{k_1}}\frac{d^3k'_1}{(2\pi)^3(1-x')2\varepsilon_{k'_1}},             
\end{eqnarray}  
where $x=\omega\cd k_1/\omega\cd p$, $x'=\omega\cd k'_1/\omega\cd p$.
The factor $(\mu^2+\vec{K}\,^2)$ is the denominator of the meson propagator,
which is expressed in terms of the relative nucleon momenta. 
In these variables $\vec{K}^2$ is given by
\begin{equation}\label{k4}                                                      
\vec{K}\,^2 = (\vec{k}\,' -\vec{k}\,)^2                                         -
(\vec{n}\cd\vec{k}\,')(\vec{n}\cd\vec{k})                                           
\frac{(\varepsilon_{k'}-\varepsilon_k)^2}
{\varepsilon_{k'}\varepsilon_k} +(\varepsilon_{k'}^2
+\varepsilon_k^2-\frac{1}{2}M^2)                           
\left|\frac{\vec{n}\cd\vec{k}\,'}{\varepsilon_{k'}}                                
-\frac{\vec{n}\cd\vec{k}}{\varepsilon_k}\right|\ .                                 
\end{equation}    

In the time ordered graph technique, the analytical expressions for the
diagrams (a) and (b) in figure \ref{ct4} differ by the meson propagators. The
modulus  in (\ref{k4}) just take into account this change of the meson
propagator for different time ordering of the vertices. Therefore 
expression (\ref{c1}) with (\ref{k4}) for $\vec{K}\,^2$ corresponds to the sum
of graphs (a) and (b).

Expression (\ref{c1}) is obtained similarly to the case  of the impulse
approximation. The factor $V$ stands for the meson-nucleon vertex at the upper
and lower lines. The factor $(-U_c\bar{\phi}'_{\mu})$ appears from  relation
(\ref{wfc}). Taking again into account the fact that $U_c\hat{k}^t=-\hat{k}U_c$ we get: 
\begin{eqnarray}\label{c2}                                                     
J^{\rho}_{\mu\nu}\vert_{left}=\frac{m}{2^7\pi^6(\omega\cd p)} \int                 
Tr\{\phi'_{\mu}                                                                 
(\hat{k}'_2+m) \Gamma^{\rho}\hat{\omega} V (\hat{k}_2+m)\phi_{\nu}              
(\hat{k}_1-m)V^c (\hat{k}'_1-m) \} \nonumber \\                                 
\frac{1}{(\mu^2+\vec{K}\,^2)} \frac{1}{(1-x')} 
\frac{d^3k} {\varepsilon_k}\frac{d^3k'}{\varepsilon_{k'}},                      
\end{eqnarray}                                                                  
where $V^{c}=U_c V^{t}U_c$.  We used the relation: $\omega \cd
k'_2=(1-x')\omega\cd p$.  Instead of momenta $\vec{k}_1,\vec{k}_2$
we integrate in (\ref{c2})  over the relative nucleon momenta
$\vec{k},\vec{k}\,'$, using the relation:                             
\begin{equation}\label{kin9}                                                    
\frac{d^3k_1}{2(1-x)\varepsilon_{k_1}}=                                         
\frac{d^3k}{\varepsilon_k}=\frac{d^2 R_{\perp}dx}{2x(1-x)}                      
\end{equation}                                                                  
and similarly for $k'_1,\vec{k}\,',\vec{R}'_{\perp}$. These relative momenta
are the arguments of the initial and final deuteron wave functions
respectively.

The right contact term is obtained from eq.(\ref{c2}) by the replacement                                                                     
\begin{equation}\label{c3}                                                      
\frac{\Gamma^{\rho}\hat{\omega} V}{(1-x')} 
\rightarrow\frac{V\hat{\omega}\Gamma^{\rho}}{(1-x)}                                       
\end{equation}                                                                  

In case of scalar (pseudoscalar) exchanges one should put both for up
and down vertices $V=V^{c}=g$ ($V=V^{c}=ig\gamma_5$). For the
pseudoscalar exchange one can simplify the contact terms by excluding $\gamma_5$: 

\begin{eqnarray}\label{c4}                                                      
J^{\rho}_{\mu\nu}|^{ps}_{left}=-\frac{m}{2^7\pi^6(\omega\cd p)}                    
\int Tr\{\phi'_{\mu}                                                            
(\hat{k}'_2+m)                                                                  
\Gamma^{\rho}\hat{\omega}                                                       
(\hat{k}_2-m)\tilde{\phi}_{\nu}                                                 
(\hat{k}_1+m)(\hat{k}'_1-m)\}\nonumber \\                                                                    
\times                                                                          
\frac{g^2}{(\mu^2+\vec{K}\,^2)}                                                 
\frac{1}{(1-x')}                                                                
\frac{d^3k} {\varepsilon_k}\frac{d^3k'}{\varepsilon_{k'}},                      
\end{eqnarray}                                                                  
\begin{eqnarray}\label{c5}                                                      
J^{\rho}_{\mu\nu}|^{ps}_{right}=-\frac{m}{2^7\pi^6(\omega\cd p)}                   
\int Tr\{\tilde{\phi}'_{\mu}                                                    
(\hat{k}'_2-m)                                                                  
\hat{\omega}                                                                    
\Gamma^{\rho}                                                                   
(\hat{k}_2+m)\phi_{\nu}                                                         
(\hat{k}_1-m)                                                                   
(\hat{k}'_1+m) \} \nonumber \\                                                  
\times                                                                          
\frac{g^2}{(\mu^2+\vec{K}\,^2)}                                                 
\frac{1}{(1-x)}                                                                 
\frac{d^3k} {\varepsilon_k}\frac{d^3k'}{\varepsilon_{k'}},                      
\end{eqnarray}                                                                  
where $\tilde{\phi}=\gamma_5 \phi \gamma_5$. The function $\tilde{\phi}$
differs from $\phi$, eq.(\ref{eq1}), by changing the sign of 
$\varphi_2$ (and also of $\varphi_4$ and $\varphi_6$ if they are  not
neglected).                

Let us now find the vertices $V$ and $V^{c}$ for the vector exchange. It is
convenient to take them  from the expression of the corresponding kernel which has the form \cite{cdkm}:                                                    
\begin{eqnarray}\label{c6}                                                      
K=\int\bar{u}'_2(g\gamma^{\alpha}-                                                    
\frac{f}{2m}\sigma^{\alpha'\alpha}i(k-\omega\tau_1)_{\alpha'}) u_2              
\left[-g_{\alpha\beta}+                                                         
\frac{(k-\omega\tau_1)_{\alpha}(k-\omega\tau_1)_{\beta}}{\mu^2}\right]          
\nonumber \\                                                                    
\times \bar{u}'_1(g\gamma^{\beta}+                                              
\frac{f}{2m}\sigma^{\beta'\beta}i(k-\omega\tau_1)_{\beta'}) u_1 \nonumber\\                                                                     
\times\delta((k_1-k'_1+\omega\tau_1-\omega\tau)^2-\mu^2)                        
\theta(\omega\cd (k_1-k'_1))\frac{d\tau_1}{\tau_1-i\epsilon} \nonumber \\           
+\int \bar{u}'_2(g\gamma^{\alpha}+                            
\frac{f}{2m}\sigma^{\alpha'\alpha}i(k-\omega\tau_1)_{\alpha'}) u_2              
\left[ -g_{\alpha\beta}+                                                        
\frac{(k-\omega\tau_1)_{\alpha}(k-\omega\tau_1)_{\beta}}{\mu^2}\right]\nonumber \\                                                                    
\times \bar{u}'_1(g\gamma^{\beta}-                                              
\frac{f}{2m}\sigma^{\beta'\beta}i(k-\omega\tau_1)_{\beta'}) u_1\nonumber\\                                                                     
\times\delta((k'_1-k_1+\omega\tau_1-\omega\tau')^2-\mu^2)\theta(\omega\cd (k'_1-k_1))\frac{d\tau_1}{\tau_1-i\epsilon},                       
\end{eqnarray}  
$k$ is the meson momentum. The coupling constant $g$ stands for the vector
vertex $\bar{u}\gamma^{\alpha}u$, whereas $f$ corresponds
to the derivative coupling. The vector meson exchange generates
its own vector contact interaction, corresponding to the cross on the vector meson line.
This is an additional contact interaction in the vertex $NN\gamma V$.
As indicated in \cite{cdkm}, the
contact term can be taken into account by the replacement, in the numerator,
of the momentum $k$ by the difference $k-\omega\tau$, where $\omega\tau$ is the
momentum of the spurion line connecting the ends of the vector meson line.
This replacement has been done in (\ref{c6}). The factor $\omega\tau_1$
everywhere in $(k-\omega\tau_1)$
incorporates the contact terms both for the vector meson exchange and for the
derivative coupling. We express $k$ in terms of the lower vertex momenta: 
\begin{eqnarray}\label{c6a}
&&k-\omega\tau_1=k_1-k'_1-\omega\tau\quad\mbox{for}\quad 
\omega\cd (k_1-k'_1) > 0  \nonumber\\
&&k-\omega\tau_1 = k'_1-k_1-\omega\tau'\quad\mbox{for}\quad 
\omega\cd (k'_1-k_1) > 0
\end{eqnarray}
The first line of (\ref{c6a}) corresponds to the first item in (\ref{c6}),                              
the second line corresponds to the second one.
                                                                                
By this way, we find the vector contact term (both the left and the right
one). It is convenient to represent it as the sum of two contributions:                      
\begin{equation}\label{c6b}
J^{\rho}_{\mu\nu}|^{vector}= J^{(1)\rho}_{\mu\nu}+                            
J^{(2)\rho}_{\mu\nu}.
\end{equation} 
                                                       
The first one, $J^{(1)\rho}_{\mu\nu}$, arises from the contraction of             
the $NN$-meson vertices with $-g_{\alpha\beta}$ in the meson propagator appearing in (\ref{c6}).
For example,  $J_{left}$ is obtained                                                       
by the following substitution in eq.(\ref{c2}) (and similarly for $J_{right}$):                                                                   

\begin{equation}\label{c7}                                                      
V = g\gamma^{\alpha}-\frac{f}{2m}\sigma^{\alpha'\alpha}i(k_1-k'_1- \omega\varphi(\tau,\tau'))_{\alpha'},                                                 
\end{equation}                                                                  
\begin{equation}\label{c8}                                                      
V^c \rightarrow g\gamma_{\alpha}+\frac{f}{2m}\sigma_{\beta'\alpha}i(k_1-k'_1-\omega\varphi(\tau,\tau'))^{\beta'}.                                                  
\end{equation}

The vertex $V^{c}$ is given by (\ref{c8}) with a minus sign; this sign is
compensated by the minus from $-g_{\alpha\beta}$, what gives  the substitution
(\ref{c8}). In eqs.(\ref{c7},\ref{c8}) we introduce the function, which takes into
account the condition (\ref{c6a}):                       

\begin{equation}\label{cp}                                                      
\varphi(\tau,\tau')=\left\{\begin{array}{rrl}                                   
\tau, &\mathrm{ if} & x>x' \\                                                       
-\tau', &\mathrm{ if} & x<x' \\                                                     
\end{array} \right.                                                             
\end{equation}                                                                  

Since the vertex $V^{c}$ appears multiplied on the left by $(\hat{k}_1-m)$ 
and on the right by  $(\hat{k}'_1-m)$, it can be transformed into:          
\begin{equation}\label{c9}                                                      
V^{c}\rightarrow(g+f)\gamma_{\alpha}-\frac{f}{2m}(k_1+k'_1)_{\alpha}-i\frac{f}{2m}\sigma_{\beta'\alpha}\omega^{\beta'}\varphi(\tau,\tau')            
\end{equation}                                                                  
                                                                                
The second item $J^{(2)\rho}_{\mu\nu}$ in (\ref{c6b}) arises from the
contraction of the vertices with the factor
$(k-\omega\tau_1)_{\alpha}(k-\omega\tau_1)_{\beta}$ in
(\ref{c6}). It is given by (\ref{c2}) with the following vertices 
\begin{equation}\label{c10}                                                     
V \rightarrow \frac{g}{\mu}(\hat{k}_1-\hat{k}'_1),\,\,\,\,                      
V^{c} \rightarrow\frac{g}{\mu} \hat{\omega}\varphi(\tau,\tau').                 
\end{equation}                                                                  
                                                                                
The other terms, like the term proportional to $\hat{\omega}$, do not
contribute to the upper vertex, since being multiplied by $\hat{\omega}$ from
the contact term it gives zero. The difference $\hat{k}_1-\hat{k}'_1$ in the
lower vertex appears in the expression 
$(\hat{k}_1-m)(\hat{k}_1-\hat{k}'_1)(\hat{k}'_1-m)=0$ and, hence,  does not
contribute as well. The minus sign  from
$U_c\gamma_{\beta}^tU_c=-\gamma_{\beta}$ is incorporated in the function
(\ref{cp}).

\section{Calculating form factors}\label{ffs}

The next step consists in extracting the deuteron form factors from the  electromagnetic
vertex $J_{\mu\nu}^{\rho}$.  As already mentioned, in  contrast to the wave
function which is always off energy shell, the on-shell amplitudes should not
depend on the light-front  plane orientation.
However, in practice, due to
the incompatibility of the transformation properties of the approximate 
current and wave function, the nonphysical $\omega$ dependence survives
in the  on-energy shell deuteron electromagnetic vertex.  Thanks to covariance, 
the general form of this dependence can be found explicitly \cite{ks9294}: 
\begin{eqnarray}\label{eq6} 
J_{\mu\nu}^{\rho}&=& 
P^{\rho}\left[ {\cal F}_1 g_{\mu\nu} +{\cal F}_2 {q_{\mu}q_{\nu}\over 
2M^2}\right] + {\cal G}_1 (g_{\mu}^{\rho}q_\nu - g_{\nu}^{\rho}q_\mu) \nonumber\\
&+& B_1{M^2\omega^{\rho} g_{\mu\nu}\over 2\omega\cd p}
+ B_2  {\omega^{\rho} q_{\mu}q_{\nu}\over 2\omega \cd p }
+ \cdots + B_8q^{\rho}\frac{q_{\mu}\omega_{\nu}+
q_{\nu}\omega_{\mu}}{2\omega\cd p}. 
\end{eqnarray}
Here $q=p'-p$, $P=p+p'$, with $p$ and $p'$  the initial and final deuteron momenta. As
an example, we keep in (\ref{eq6}) three $\omega$-dependent terms only, though
the total number of them is eight. Since we assume  $\omega\cd q=0$ (what
corresponds to $q_+=0$ in the standard approach), all the form factors in
(\ref{eq6}) depend on $Q^2=-q^2$ only.

The $\omega$-dependence of the wave function is not the only source of the
nonphysical contributions $B_{1-8}$.  With an $\vec{n}$-independent wave function
(neglecting $f_5$ in (\ref{eq2})) we still get the $\omega$-dependent 
deuteron electromagnetic vertex. On the contrary, the physical form  factors
${\cal F}_1,{\cal F}_2,{\cal G}_1$ do not depend on $\omega$ even with an 
$\omega$-dependent wave function.  We emphasize that the physical 
$\omega$-dependent extra components in the wave function (\ref{eq2}) and the
nonphysical $\omega$-depending structures in (\ref{eq6}) are  present not
because of the covariant formulation of the LFD.  Their counterparts appear in
the non-covariant light-front approach too but the covariant approach allows to
indicate them explicitly.

The explicit formulas to extract the physical form factors ${\cal F}_1, {\cal
F}_2, {\cal G}_1$ from the electromagnetic vertex and to
separate them from  nonphysical contributions $B_{1-8}$  were derived in
refs. \cite{ks9294}. As an example, we give here the
formula for ${\cal F}_1$ only:
\begin{equation}\label{eq7} 
{\cal F}_1 = J_{\mu\nu}^{\rho}{\omega_{\rho} \over{2\omega\cd p}} 
\left[g^{\mu\nu} - {q^{\mu}q^{\nu} \over{q^2}} - {{P^{\mu}\omega^{\nu} 
+ P^{\nu}\omega^{\mu}} \over{2\omega\cd p}}+ 
P^2{{\omega^{\mu}\omega^{\nu}} \over{4(\omega \cd p)^2}}\right],
\end{equation}
The contraction with $\omega_{\rho}$ in (\ref{eq7}) corresponds
in the  standard approach to the component $J_{\mu\nu}^+$. The expression for
${\cal F}_2$ is proportional to the same contraction. However the expression
for ${\cal G}_1$ contains not only the contraction with $\omega_{\rho}$ but
also others terms. Hence, the $J^+$ component is not enough to find ${\cal G}_1$.

One can easily check, using $\omega\cd q=0$, that one has for the nucleon electromagnetic vertex (\ref{eq5})  
$\hat{\omega}\omega_{\rho}\Gamma^{\rho}=\omega_{\rho}\Gamma^{\rho}\hat{\omega}=0$.
Here $\hat{\omega}$ appears from
the contact interaction, the  contraction $\omega_{\rho}\Gamma^{\rho}$ is the
result of eq.(\ref{eq7}). 
Therefore  the contact
terms  do  not contribute to ${\cal F}_1,{\cal F}_2$, due to the  contraction with
$\omega_{\rho}$  in (\ref{eq7}).  They contribute only to ${\cal G}_1$. The
form factors found this way are still not exact, but they are separated 
from the non-physical ones $B_{1-8}$. As shown in \cite{ks9294}  and in
\cite{karm96}, this separation is important, since the admixture of $B_i$ can
considerably change the results.

After calculating traces and contracting tensors in eq.(\ref{eq7})
for ${\cal F}_1$ (and similar expressions \cite{cdkm}  for ${\cal F}_2$ and
for ${\cal G}_1$) we obtain the integrand depending on the scalar products of
the four-vectors $\omega$, $k_1$, $k_2$, $k'_1$, $k'_2$, $p$, $p'$,  $q$ with
each other. In their turn, these scalar products, as  well as the arguments of
the wave functions, are expressed through the integration momenta and the
momentum transfer.  Their explicit  kinematical formulas are given in
\cite{cdkm}.  The integrands  also contain the scalar products with $\omega$ but
after integration this dependence disappears.  

By this way, substituting $J_{\mu\nu}^{\rho}$ in
eq.(\ref{eq7}) (and in similar expressions \cite{ks9294} for ${\cal F}_2,{\cal G}_1$),
we find the analytical expressions of the form factors integrands. Then we carry out the
numerical integration which, for the contact terms (two-loop), is 6-dimensional.  

\section{Influence of the nucleon EM form factors}\label{sec_nemff}

\begin{figure}[htbp]
\epsfxsize=15cm\epsfysize=18cm\centerline{\epsfbox{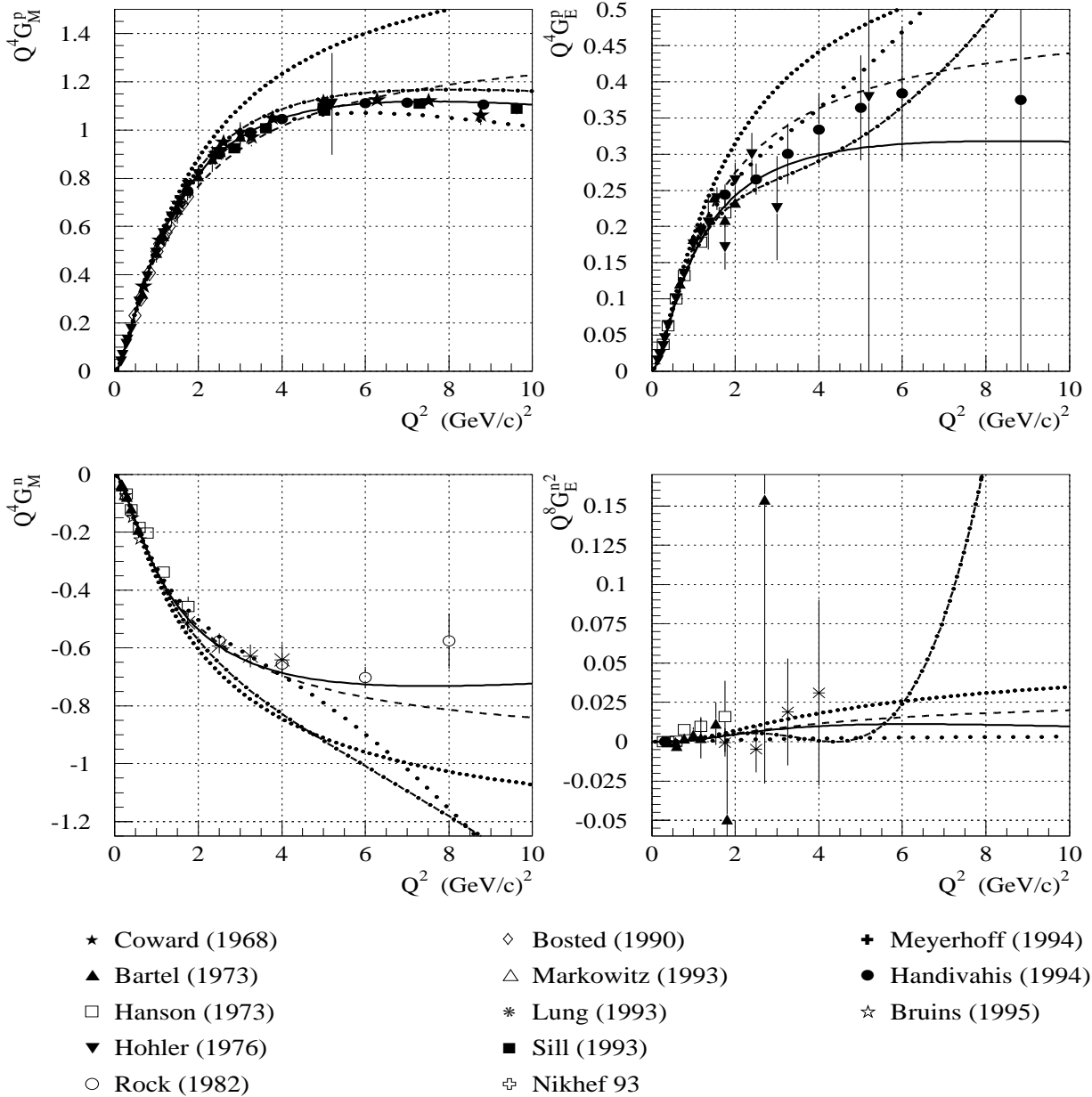}}
\caption{
Comparison  between proton ($G^p_{E,M}$) and neutron ($G^n_{E,M}$)
 electromagnetic form factors parametrizations 
and experimental data taken from \protect{\cite{exp_nemff}}.
The continuous line corresponds to MMD \protect{\cite{MMD_96}},
dot-dashed (-$\cdot$-$\cdot$-) to H\"ohler et al. \protect{\cite{HPSBSWW_76}},
dashed (-- -- --) to Galster et al. \protect{\cite{GKMSWB_71}},
short-dotted (\ldots) to BKL      \protect{\cite{BKL_71}} and 
long-dotted  ($.\;\;.\;\;.$) to SP \protect{\cite{SP}    }         
 }\label{exp_nemff1}
\end{figure}

Together with the deuteron wave function, an essential ingredient in these
calculations are the nucleon  electromagnetic form factors  (NEMFF). The
results presented in \cite{cdkm} were obtained with the dipole parametrization
taken from Bilenkaya et al. \cite{BKL_71} denoted hereafter by BKL.  We analyse
in what follows the influence of different NEMFF parametrizations  in
calculating the deuteron structure function within the same theoretical scheme.
The parametrizations we consider, used in similar calculations found in the
literature \cite{AAD_98,HT_89,VDG_95,SR_91,AV18_95},
are those given by Galster et al.  \cite{GKMSWB_71}, Hohler et al.
\cite{HPSBSWW_76},  a combination of Simon et al. for proton and Platchkov et
al. for neutron (SP) \cite{SP} and  the more recent publication of Mergell et
al. (MMD) \cite{MMD_96}. The comparison \cite{Real} of the proton ($G^p_E,G^p_M$)  and
neutron ($G^n_E,G^n_M$) charge and magnetic form factors  with the existing
experimental data \cite{exp_nemff} is shown in figure \ref{exp_nemff1} . The
values of the neutron charge form factor have been squared to account for the
results reported in \cite{exp_nemff}. A first sight to this figure  shows that
some of these form factors have been apparently determined in the low momentum
region and can not be used above 1-2 (GeV/c)$^2$. The more adequated
parametrization covering the whole momentum region seems to be MMD
\cite{MMD_96}, which will be adopted in our calculations. It provides an
acceptable description for all the form factors although some variations can
not be excluded due to the inaccuracy of the existing measurements, especially
for the neutron form factor (see below). Any comparison between the theoretical
and experimental results has to be understood in the context of this
uncertainty that we would like to estimate.

\begin{figure}[htb]
\begin{center}
\hspace{.5cm}
\subfigure[]{\epsfxsize=7.8cm\epsfysize=9cm\mbox{\epsffile{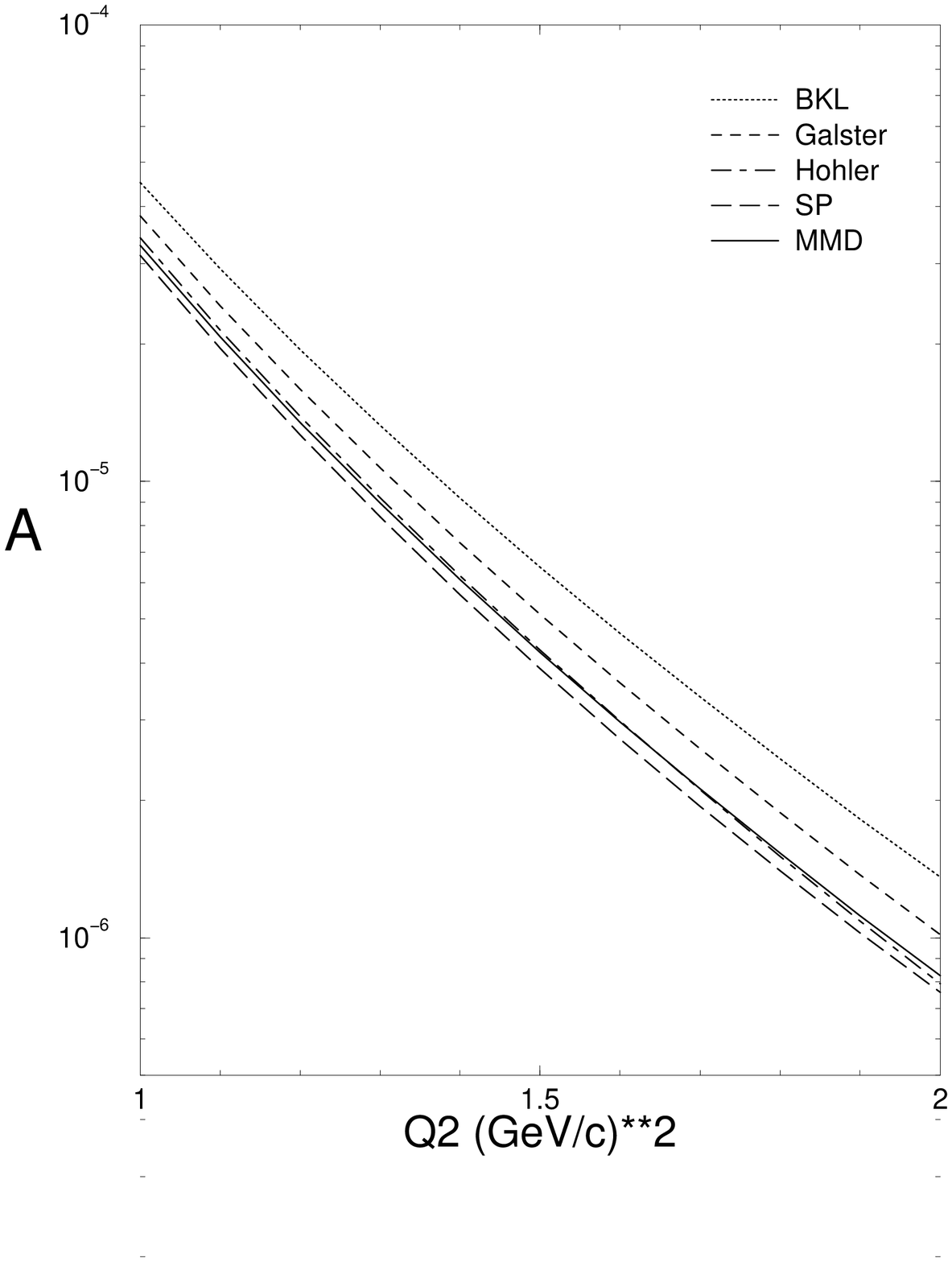}}}
\hspace{.5cm}
\subfigure[]{\epsfxsize=7.8cm\epsfysize=9cm\mbox{\epsffile{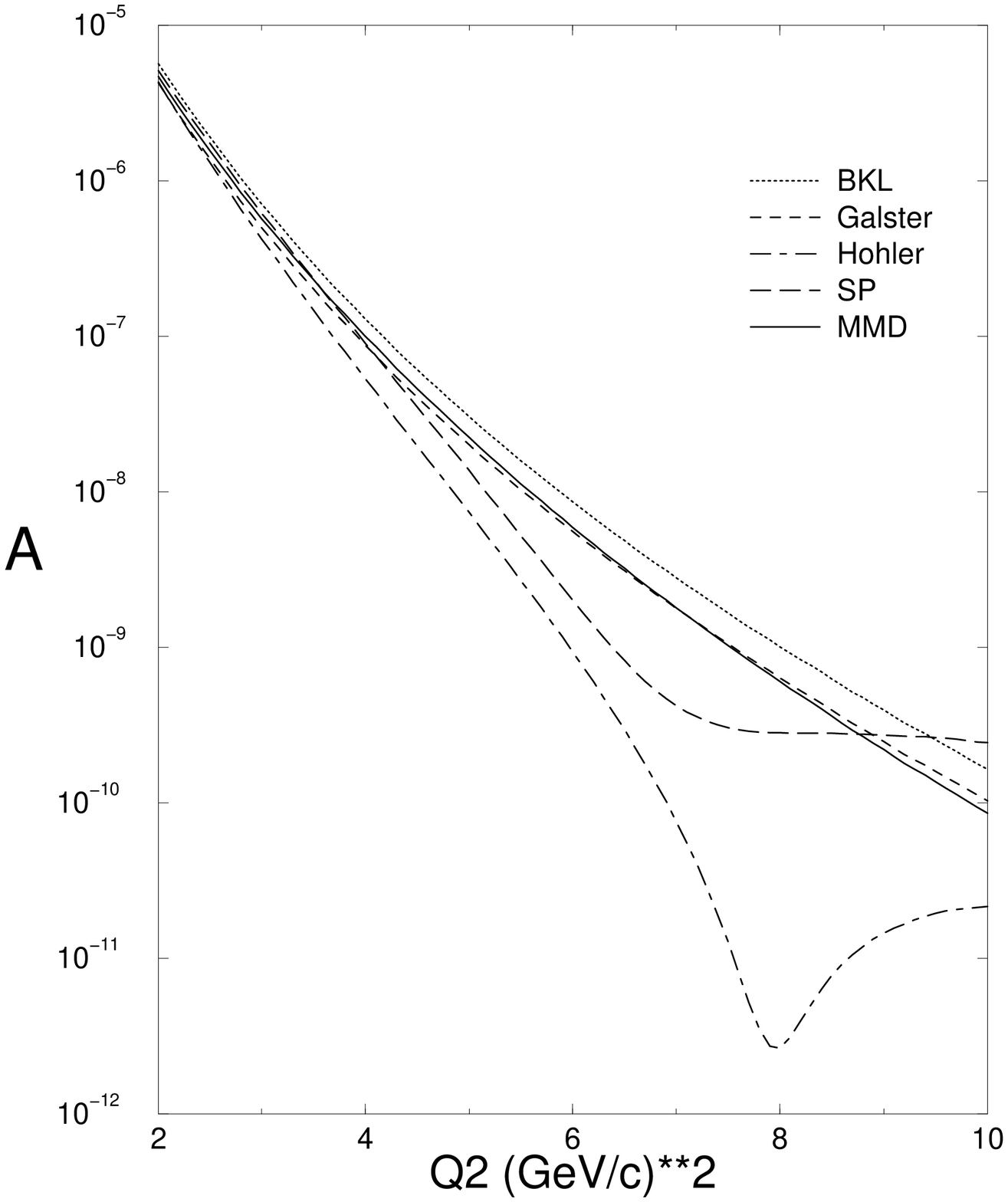}}}
\end{center}
\caption{Influence of the nucleon electromagnetic form factors in the  deuteron structure function $A(Q^2)$ 
 in the non relativistic impulse approximation (a) and in the LFD impulse approximation (b)}\label{A_EMFF}
\end{figure}

The influence of different NEMFF parametrizations  on the deuteron  structure
function $A(Q^2)$ is shown in figure \ref{A_EMFF}. The results in  figure
\ref{A_EMFF}(a) are fully non relativistic whereas those in  \ref{A_EMFF}(b)
correspond to LFD impulse approximation.  The choice of the NEMFF has small
influence at low momentum transfer. For instance at $Q^2=0.5$ (GeV/c)$^2$,  the
deviation in the  nonrelativistic structure function $A$ is $\approx 15\%$.
This deviation increases with $Q^2$ and reaches already a factor two at $Q^2=2$
(GeV/c)$^2$ between BKL and SP parametrizations. The difference becomes
dramatic for the last measured values $Q^2=6$ (GeV/c)$^2$ where there is one
order of magnitude in the structure functions $A(Q^2)$  calculated using BKL or
Hohler parametrizations.   It is worth noticing however that the Galster and
MMD sets lead to close  results in all the momentum range. Their difference in
the LFD-IA relativistic calculation of $A$ remains at the level of 10 \% for
$Q^2\leq 10$ (GeV/c)$^2$. Similar deviations have been found in the structure
function $B$.

As one can see from figure \ref{exp_nemff1}, the neutron electric form factor
$G^n_E$ contains the greatest uncertainty. To estimate the influence of this
uncertainty,  we have considered for $Q^2=2-4$ (GeV/c)$^2$ the maximal
${G_E^n}^2$  values compatible with the error bars, which are a priori not
excluded by the experimental data although they considerably differ from all
fits.  The values of $A(Q^2)$ obtained with such a modification exceed by
20-30\%  those obtained with the original MMD parametrization. From these
considerations it follows that  {\it i}) the comparison between different
calculations is only meaningful when the same parametrization of NEMFF is used
and  {\it ii}) the comparison between a theoretical prediction and the measured
deuteron form factors is limited by the considerable uncertainties implied by
the poor knowledge of the NEMFF for $Q^2>1$ (GeV/c)$^2$.

\section{Results and discussion}\label{numer}

We present in this section the comparison of the deuteron structure functions
and tensor polarization observable with the last measured values. In terms of
the deuteron form factors these observables  read
\begin{eqnarray}                                         
A(q^2)&=&F^2_{C}(q^2)+\frac{8}{9}\eta^2 F^2_Q(q^2)+\frac{2}{3}                           
\eta F^2_{M}(q^2)\ , 
\label{sfa}\\                                                 
B(q^2)&=&\frac{4}{3}\eta(1+\eta) F^2_{M}(q^2)\ .  \label{sfb}                                 
\end{eqnarray}      
\begin{equation}\label{t20}
t_{20}\left(A(q^2)+\tan^2\frac{1}{2}\theta\; B(q^2)\right)=
-\frac{1}{\sqrt{2}}\left[\frac{8}{3}\eta F_C F_Q +\frac{8}{9}\eta^2 
F_Q^2 +\frac{1}{3}\eta\left(1+2(1+\eta)  
\tan^2\frac{1}{2}\theta\right)F_M^2\right],
\end{equation}
where 
\begin{eqnarray*}                                                      
F_{C} &=& - {\cal F}_1 -{ 2\eta \over{3}}[{\cal F}_1 + {\cal G}_1 - {\cal F}_2(1 + \eta)]\ ,\\                                                                                                                      
F_{M} &=& {\cal G}_1\ ,\\                                                                                  
F_Q   &=& -{\cal F}_1 - {\cal G}_1 + {\cal F}_2(1 + \eta)\ ,                      
\end{eqnarray*}                                                                  
and $\eta = Q^2/4M^2$.
In comparing our calculations with the data we should first emphasize that,
though the theoretical framework we use -- the light-front dynamics -- is fully
selfconsistent, the validity of our perturbative  method to evaluate the wave
function is restricted to relative nucleon momenta $k$ smaller than the nucleon
mass $m$. Since in a form factor calculation the momentum transfer is
distributed between two deuteron vertices, the initial and final ones, this value of
$k$ corresponds approximately to the momentum transfer $Q^2\leq(2m)^2\approx
3.5$  (GeV/c)$^2$. In absence of subtle cancellations increasing the
uncertainty in the theoretical predictions, we can expect  a reasonable
description of the experimental data in this momentum region.

\begin{figure}[htbp] 
\hspace{0.8cm}\mbox{\epsfxsize=7.7cm\epsfysize=9cm\epsffile{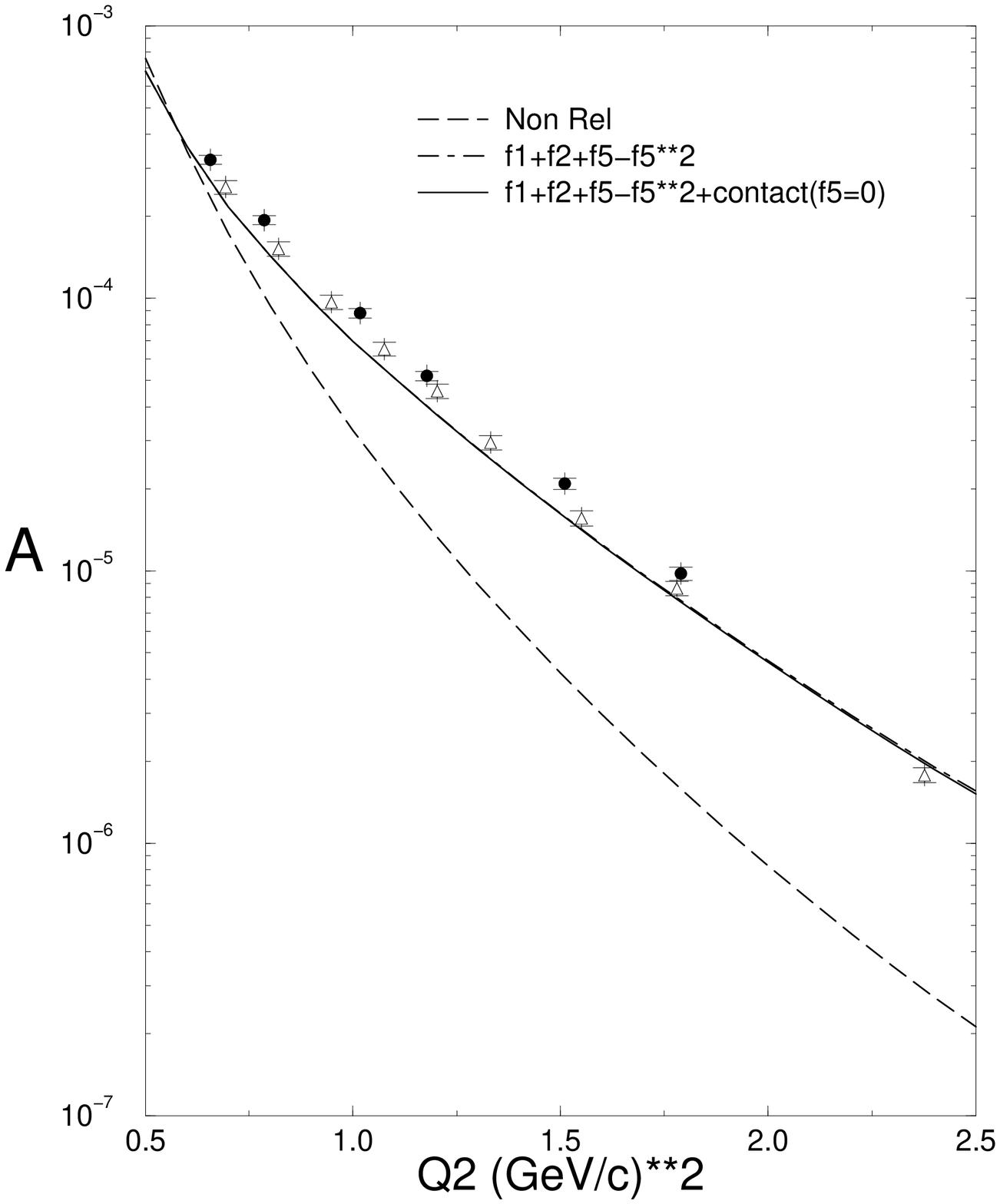}\hspace{0.5cm}
                    \epsfxsize=7.7cm\epsfysize=9cm\epsffile{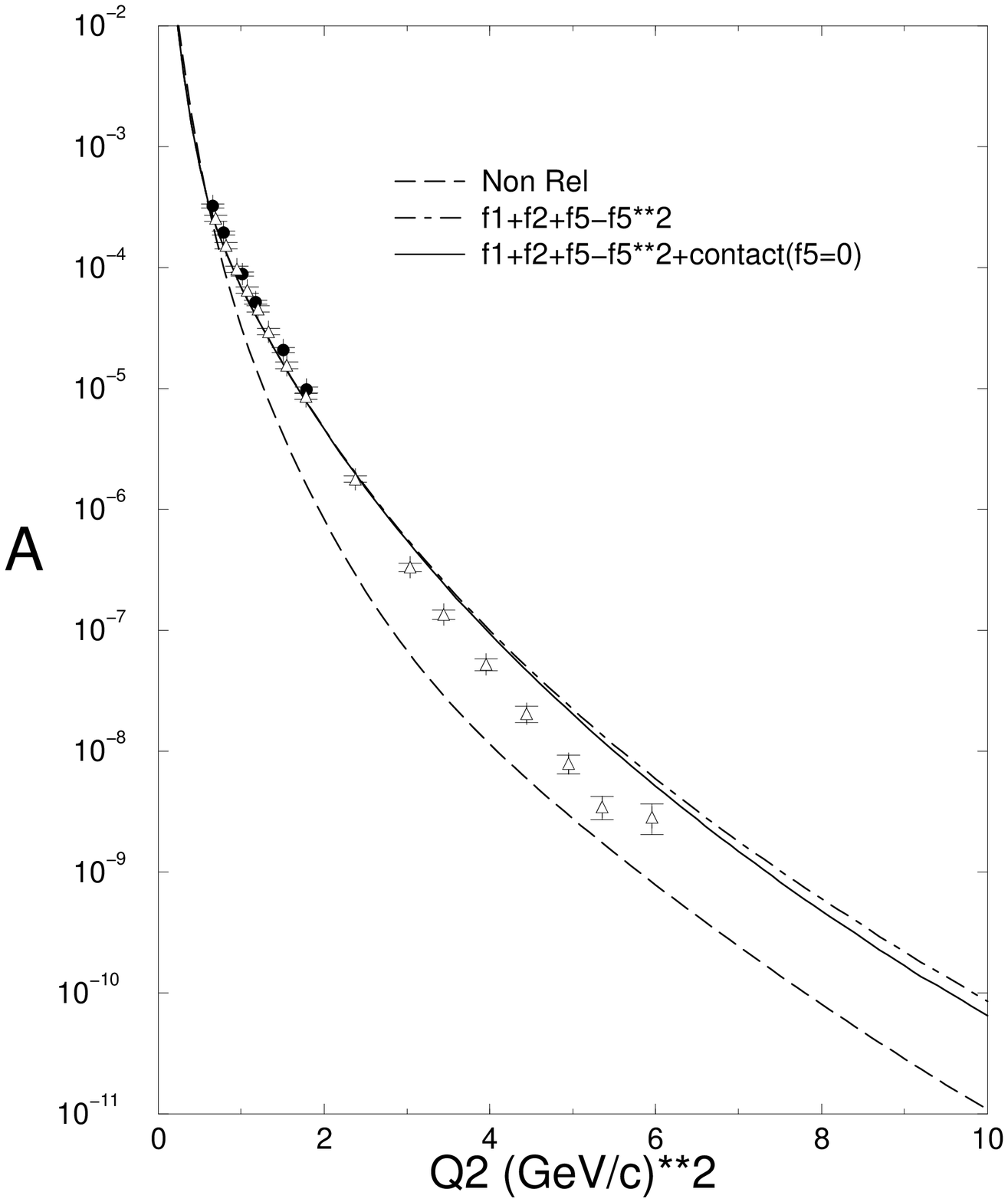}} 
\caption{The structure
function $A(Q^2)$ of the deuteron with MMD nucleon form factors.  The curves
are explained in the text.}\label{A1} 
\end{figure}

In figure \ref{A1} the structure function $A(Q^2)$ is shown together with the 
experimental points recently obtained at TJNAF (Hall C in filled circles
\cite{Kox} and Hall A in opaque triangles up \cite{Gomez}). The dashed  curve
corresponds to the non-relativistic impulse approximation with  the S- and
D-waves of the Bonn-QA wave function  \cite{bonn}. The dot-dashed line is also
calculated in the impulse approximation but using the light-front  formalism 
with  relativistic deuteron components
$f_1$ and $f_2$ and  the  $f_5$ component in first degree only. The solid line
incorporates,  in addition, the contact terms which turn to have small influence
in this observable. We can see a  good agreement until $Q^2=2.5$ (GeV/c)$^2$,
a momentum region where the departure from a non relativistic
description reaches one order of magnitude.
It is worth noticing that this agreement is found
with the LFD wave function alone, i.e. without explicitly including  any MEC
diagrams. However, as it has been already mentioned, the extra component $f_5$ accounts for the so
called pair terms in the deuteron electrodisintegration amplitude \cite{dkm95}. 
We remark that a systematic deviation seems to manifest above
3 (GeV/c)$^2$. This excess is smoothly increasing with increase of $Q^2$. 
However, this is the region where the perturbative  calculation is hardly
reliable, since the deuteron wave function appears in $A(Q^2)$ in the fourth
degree and a 50\% correction may change the value by a factor 5. 
Besides, as it has been shown above, at $Q^2>2$ (GeV/c)$^2$ the
uncertainty coming from the NEMFF is also high.

\begin{figure}[htbp]
\epsfxsize=8cm\epsfysize=9cm\centerline{\epsfbox{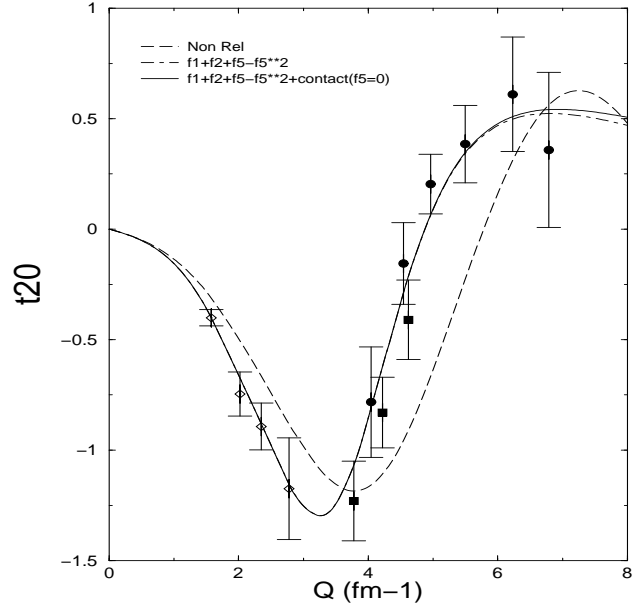}}
\caption{The deuteron tensor polarization $t_{20}$ at 
$\theta=70^{\circ}$.  The designations of the curves are the same as in 
figs.  \protect{\ref{A1}}.}\label{exp_t20}
\end{figure}

The same agreement is seen in the deuteron 
polarization observable $t_{20}$ displayed in figure \ref{exp_t20} using the same drawing
conventions. Again a sizeable effect of the relativistic
corrections is properly taken into account in our
calculations although the actual error bars are quite comfortable.

\begin{figure}[htbp]
\epsfxsize=8cm\epsfysize=8cm\centerline{\epsfbox{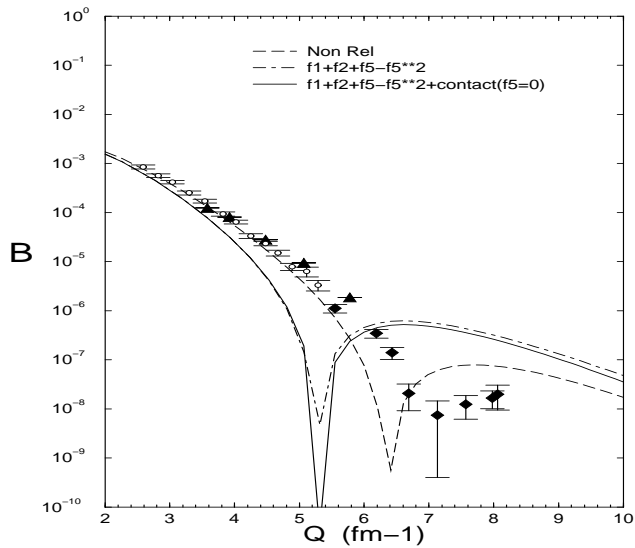}}
\caption{Same as figure \protect{\ref{A1}} but for $B(Q^2)$.}
\label{B1}
\end{figure}

The calculated structure function $B(Q^2)$ is shown in figure \ref{B1},
using the same drawing conventions,  
together with the experimental  data from \cite{exp_BQ2}.
Comparing the $B(Q^2)$ results with those of $A(Q^2)$ and
$t_{20}$, one can see a considerable deviation of our $B$ calculation from the
data, especially in the region of the minimum, as well as highest
sensitivity to the different approximations. This minimum corresponds to the 
zero of the deuteron magnetic form factor $F_M$, and a small shift on this
value drastically changes the value of $B(Q^2)$ in its neighborhood. It is
important to emphasize that the zero of $F_M$ which exists in a nonrelativistic 
calculation, {\em disappears} for the relativistic $f_1,f_2$ and {\em appears 
again} when $f_5$ is taken into account \cite{cdkm}. In a relativistic
framework this minimum is thus a consequence of a delicate  cancellation
between the $f_5$ contribution with the contributions of $f_1$ and $f_2$. None
of these  $f_1,f_2,f_5$ components has been calculated with enough accuracy in
our perturbative approach. Therefore we cannot pretend to a detailed
description of $B(Q^2)$ in this region.

\begin{figure}[htb]
\begin{center}
\hspace{.5cm}
\subfigure[$A(Q^2)$]{\epsfxsize=7.cm\epsfysize=7cm\mbox{\epsffile{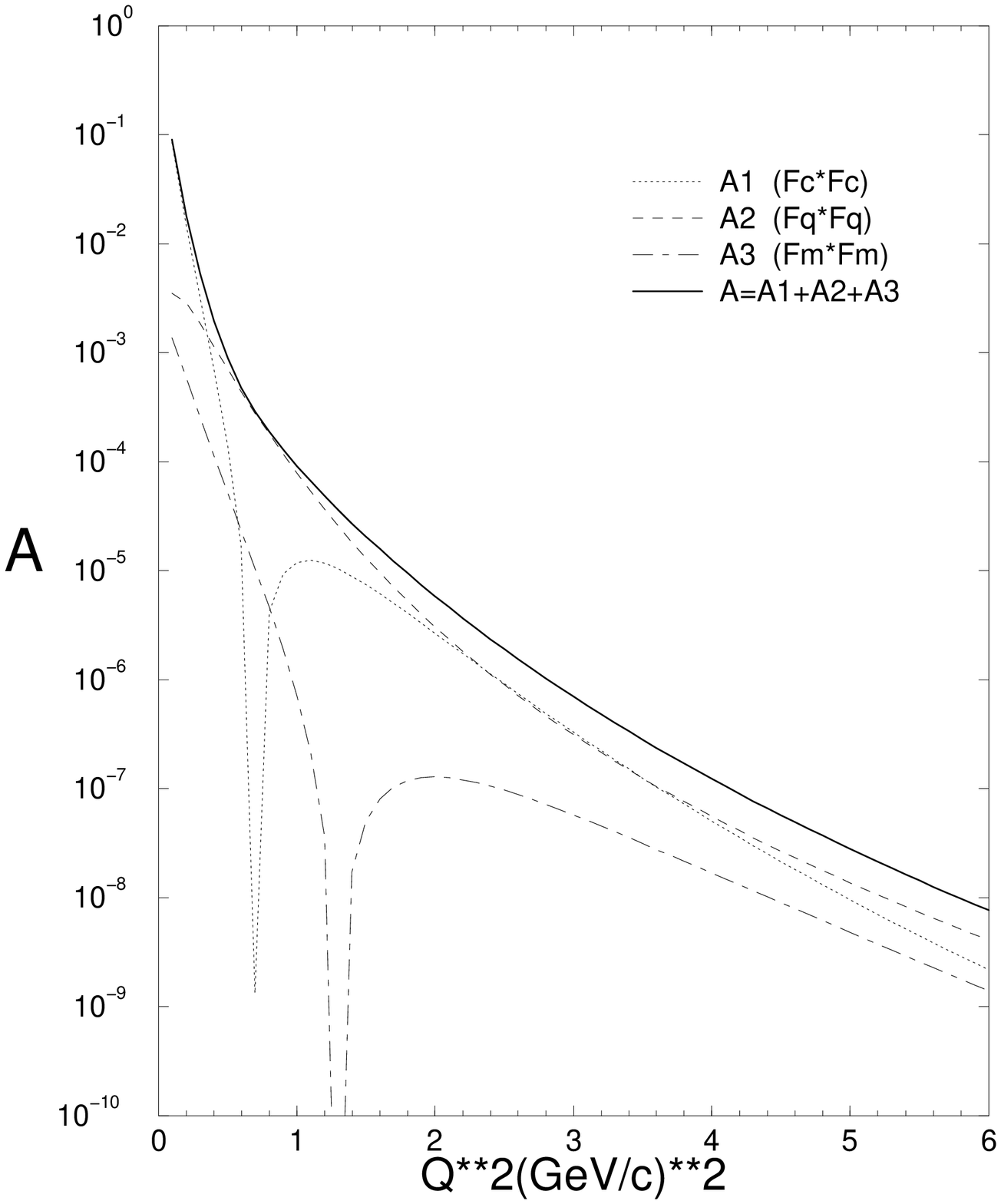}}}
\hspace{.5cm}
\subfigure[$t_{20}$]{\epsfxsize=7.cm\epsfysize=7cm\mbox{\epsffile{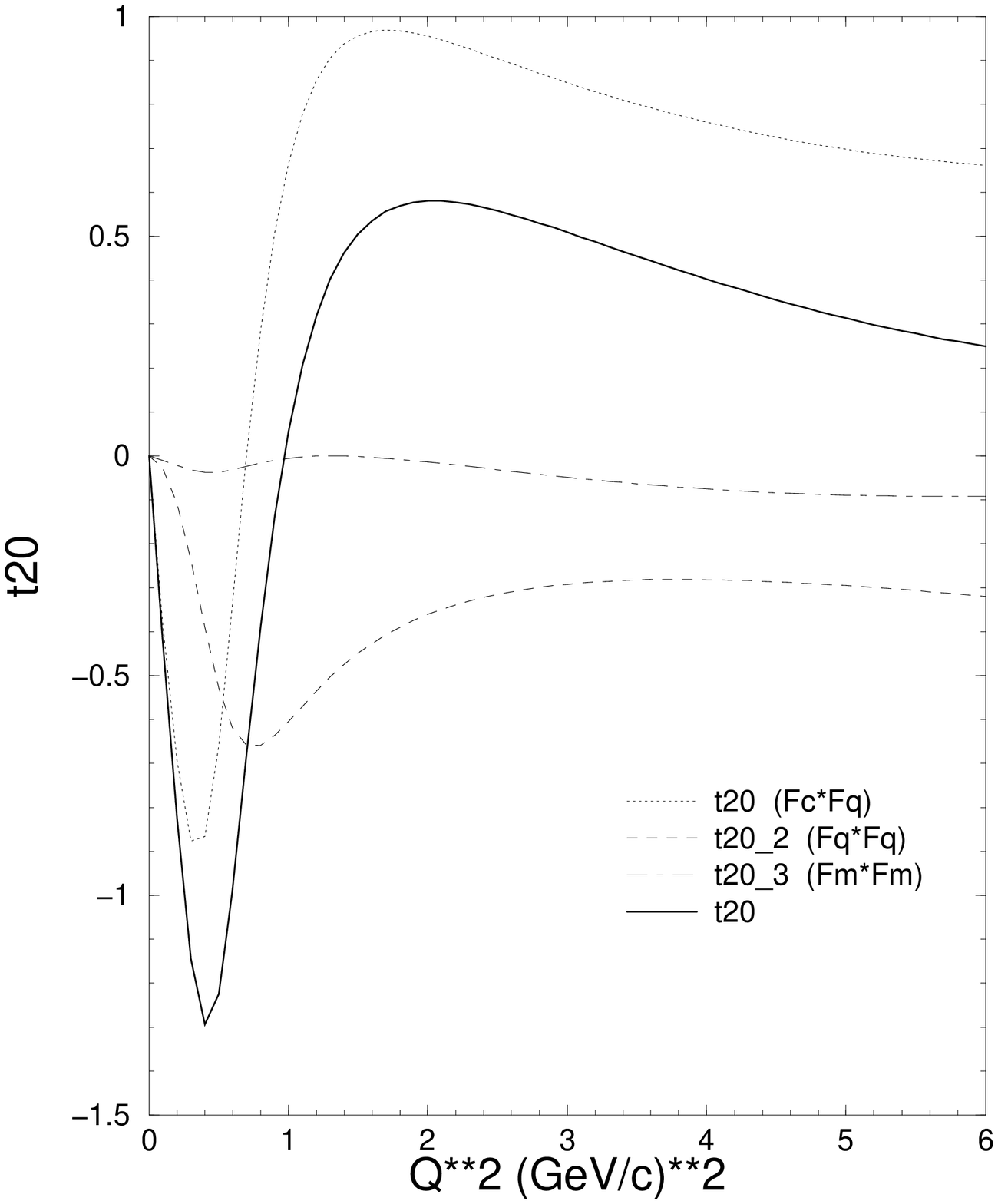}}}
\end{center}
\caption{The contributions of the different deuteron form factors 
($F_C,F_Q,F_M$) to the 
structure function $A(Q^2)$ (a) and $t_{20}$ (b).}\label{Contrib}
\end{figure}

The situation is different for $A(Q^2)$ and $t_{20}$. Figure \ref{Contrib} 
shows the relative contributions of the three deuteron form  factors to
these observables according to equations (\ref{sfa}) and (\ref{t20}).
One can see that their values  are dominated by the
charge and quadrupole form factors $F_C$ and $F_Q$, whereas the contribution of
$F_M$, containing the highest uncertainty, is suppressed. 

Taking into account the same order of accuracy, i.e.,  $g^2$, which was  kept in the
calculation of the wave function, we get a good description of the data on
$t_{20}$ as well as of $A(Q^2)$ in the same region of the momentum transfer, up
to 2 (GeV/c)$^2$. The relativistic effects in $f_1,f_2$ and the contribution
of $f_5$ are important to achieve this good description. The contribution of
the contact term to $A(Q^2)$ and $t_{20}$ is small, since, as noted above, it
contributes to $F_M$ only, whereas for $B$ it is small in the momentum transfer
region shown in figure \ref{B1}  but increases at higher momenta (e.g. by a factor 2 at
$Q^2=6$ (GeV/c)$^2$ ). These results show that the relativistic effects in the
deuteron wave function, including the extra component $f_5$, and in the
deuteron  electromagnetic vertex make considerable influence on the deuteron
form factors. Note that the effect of incorporating $f_5$ on
$t_{20}$ is  qualitatively similar to that obtained when adding the
contribution of  the pair current in non-relativistic calculations
\cite{SR_91}. 

We should finally emphasize that the results presented in this section
have been obtained without fitting any new set of parameters, neither
in the nucleon-meson form factors nor in the NN interaction kernel, which was
taken as it is given in ref. \cite{bonn}.

The agreement of our calculations with the experimental data,
as well as with some other calculations carried out in the framework of meson-nucleon dynamics 
(see e.g. \cite{GK_84}-\cite{SR_91}),
shows that the deuteron structure at small distances is understood rather well
within this theoretical framework. 
It is a remarkable fact that probing distances of the order of $0.1$ fm,
at which quarks effects should  manifest themselves in their full glory,
could be accounted by the relativistic
nucleon-meson dynamics and the phenomenological nucleon form factors.

The parameters of the effective nucleon-meson Lagrangian, fixed by
fitting the $NN$ experimental data, should be in principle
derived from QCD, what is a separate problem.
This allows us to make the following general conclusion: {\em at least in what
concerns the deuteron, the relativistic nuclear dynamics can be developed
independently of its derivation from QCD}. 

Our calculations can be improved in many  respects. An exact solution of the
equation for the deuteron is necessary and is in progress. Besides the
non-perturbative  calculation, it implies the determination of the $NN$  kernel
parameters  in the framework of the light-front equations. This  requires also
a careful treatment of higher order contributions to the kernel. Finally, the
calculation of electromagnetic observables  should include higher Fock states
($NN\pi$, etc.) and the meson exchange currents which are not included in the
wave function components, like those corresponding to the interaction of the
photon with the intermediate mesons ($\rho\pi\gamma,\ldots$).

The improvement of the experimental data on the nucleon EM form factors is
also an urgent task.

\section*{Acknowledgements}
The authors are sincerely grateful to J.S. Real for providing the
experimental compilation of the nucleon electromagnetic form factors, to B.
Desplanques for stimulating critics and valuable remarks and
to M. Mangin-Brinet for helpful discusions.


\end{document}